\documentclass[aps,prb,twocolumn,groupedaddress]{revtex4}
\usepackage{graphicx}

\begin{document}
\newcommand{\be}{\begin{equation}}
\newcommand{\ee}{\end{equation}}
\newcommand{\bea}{\begin{eqnarray}}
\newcommand{\eea}{\end{eqnarray}}
\newcommand{\nt}{\narrowtext}
\newcommand{\wt}{\widetext}

\title{Composite Dirac fermions in graphene}

\author{D. V. Khveshchenko}

\affiliation{Department of Physics and Astronomy, University of North Carolina,
Chapel Hill, NC 27599}

\begin{abstract}
Generalizing the notion of composite fermions to the case of "pseudo-relativistic" Quantum Hall phenomena in
graphene, we discuss a possible emergence of compressible states at the filling factors $\nu=\pm 1/2,\pm 3/2$.
This analysis is further extended to the nearby incompressible states viewed as Integer Quantum Hall Effect of
composite Dirac fermions, as well as those that might occur at $\nu=0,\pm 1$ as a result of (pseudo)spin-singlet
pairing between the latter.
\end{abstract}
\maketitle

The traditional interest in Quantum Hall Effect has been rekindled by the recent experiments on mono- and
double-layers of graphene where the interplay between unscreened Coulomb interactions and pseudo-relativistic
kinematics of the Dirac quasiparticles has long been expected to harbor a host of novel phenomena
\cite{semenoff}.

In graphene mono-layers, the Integer Quantum Hall Effect (IQHE) plateaus were found at the integer values
$\sigma_{xy}=\nu=(4n+2)$, $n=0,\pm 1,\pm 2,\dots $ \cite{exp} (hereafter, we put $\hbar=e=c=1$, and measure all the conductivities in units of $e^2/h\equiv 1/2\pi$). Elaborating on the earlier insight of
Refs.\cite{semenoff}, this observation was readily explained \cite{th} by treating the low-energy excitations in
graphene as (pseudo)relativistic Dirac fermions with linear dispersion and speed $v_F\sim 10^6m/s$. These
quasiparticles carry a physical spin $s=1/2$ and possess an additional orbital ("pseudo-spin" or "valley")
quantum number (hereafter referred to as $R$ and $L$) corresponding to the double degeneracy of the electronic
Bloch states in graphene.

The resulting $SU(4)$ symmetry of the non-interacting Hamiltonian survives the long-range Coulomb interactions,
although it gets broken in the presence of the Zeeman and various additional short-range (Hubbard-like) interaction terms. A number of
implications of this symmetry which generalizes independent rotations in the spin and valley subspaces have been explored in several recent papers \cite{fisher}.

For one, by drawing a parallel with the previous studies of spin-unpolarized double-layer Quantum Hall systems,
it was argued that, apart from the Dirac kinematics, the situation in graphene is similar to that occurring in
the double-layer systems in the limit of vanishing inter-layer tunneling. Thus, a graphene analog of the
Quantum Hall Ferromagnet was predicted to occur, which type of strongly correlated states would manifest itself
as additional (interaction-induced) plateaus at $all$ the integer filling factors \cite{fisher}.

In a recent experiment, additional plateaus were indeed observed at $\sigma_{xy}=0,\pm 1,\dots$ \cite{kim}, thus
suggesting a complete lifting of the spin and valley degeneracies at the lowest ($n=0$) relativistic Landau
level (LL). As an alternative interpretation, it was pointed out \cite{gusynin} that the behavior reported in
Ref.\cite{kim} can also be explained by invoking the "magnetic catalysis" scenario of Ref.\cite{dvk} where,
contrary to the predictions of Refs.\cite{fisher}, the valley degeneracy gets lifted $only$ at the $0^{th}$ LL,
in agreement with the data of Ref.\cite{kim}.

Still awaiting its observation, however, is a graphene counterpart of the Fractional Quantum Hall Effect (FQHE).
By analogy with FQHE in the conventional ("non-relativistic") two-dimensional electron gas (2DEG) with parabolic quasiparticle dispersion, one
might expect that its graphene analog can also be studied by adapting the idea of statistical flux attachement
to the case of the Dirac fermions.

In the present Letter, we discuss such a procedure, thereby setting the
stage for a systematic analysis of the (potentially, much richer than in the
case of the conventional 2DEG) realm of FQHE phenomena in graphene.

The flux attachment recipe would usually be applied to a half-(or, more generally, $1/2p$-, where $p$ is an
integer) filled uppermost LL, while the rest of the system would be treated as an inert incompressible
background. In a generic $SU(N)$-invariant system, the Chern-Simons Lagrangian implementing a transformation
from the original electrons to $N$-component composite Dirac fermions (CDFs) takes the form
$$
L=\sum_{\alpha}^N\int_{\bf r}\Psi^\dagger_{\alpha}{\hat \gamma}_\alpha
(i\partial^i+a^i_\alpha+A^i)\Psi_{\alpha}
+{1\over 4\pi}\sum_{\alpha,\beta}^N\int_{\bf r}K^{-1}_{\alpha\beta}\epsilon_{ijk}a^i_\alpha\partial^j
a^k_\beta
$$
\be +{v_F\over 4\pi}\sum_{\alpha,\beta}^N\int_{\bf r}\int_{\bf r^\prime}{ \Psi^\dagger}_{\alpha}({\bf
r}^\prime)\Psi_{\alpha}({\bf r}^\prime) {g\over {|{\bf r}-{\bf r}^\prime|}} {\Psi^\dagger}_{\beta}({\bf
r})\Psi_{\beta}({\bf r}) \ee Here $g_0=2\pi e^2/\epsilon_0v_F\sim 3$ is the bare Coulomb coupling, the vector
potential $A^i=(0,-By/2,Bx/2)$ represents an external magnetic field, and the matrices ${\hat \gamma}_\alpha=({\bf
1},{\hat \sigma}_x,(-1)^\alpha{\hat \sigma}_y)$ act in the space of spinors $\Psi_\alpha$ composed of the values of the
CDF wave functions on the two sublattices of the bi-partite lattice of graphene.

In a multi-component system, the statistical flux provided by the Chern-Simons fields $a^i_\alpha$ can be
attached in a number of different ways, the choice between which should ultimately be determined by the nature
of the ground state in question. Accordingly, there exist different choices of the integer-valued matrix ${\hat
K}$, the only condition imposed upon which is that the transformed CDFs retain their (mutual) fermionic
statistics. This requirement can be readily satisfied, provided that all the matrix elements of $\hat K$ are
even integers, though.

By varying Eq.(1) with respect to the Lagrange multipliers $a^0_\alpha$, one obtains a set of constraints \be
\rho_\alpha=<\Psi^\dagger_{\alpha}\Psi_{\alpha}>= {1\over 2\pi}\sum_\beta^NK^{-1}_{\alpha\beta}<{\bf
\nabla}\times{\bf a}_{\beta}>\ee which determine the average values of the effective fields $b_\alpha=B-<{\bf
\nabla}\times{\bf a}_\alpha>$ experienced by the CDF $\alpha$-species.

In the FQHE states viewed as the CDF IQHE, each of the CDF species occupies an integer number
$\nu_\alpha=2\pi\rho_\alpha/b_\alpha=m_\alpha$ of the effective LLs. The total electronic filling factor is then
given by the expression \be \nu=\sum^N_\alpha{2\pi\rho_\alpha\over B}=Tr({\bf 1}+{\hat K}{\hat m})^{-1}{\hat m}
\ee where ${\hat m}=diag[m_1,\dots, m_N]$.

Integrating the CDFs out in the standard manner, one obtains a quadratic Lagrangian for the vector fields \bea
L_{eff}[a_\alpha, A]={1\over 2}\sum^N_\alpha(a^i_\alpha+A^i)\Pi_{ij}^\alpha(a_\alpha^j+A^j)\nonumber\\
+{1\over 4\pi}\sum_{\alpha,\beta}^N \epsilon_{ijk}K^{-1}_{\alpha\beta}a^i_\alpha\partial^ja^k_\beta \eea where
$\Pi^\alpha_{ij}(\omega,{\bf q})$ is the CDF polarization operator.

Next, by eliminating all the statistical fields, one derives the RPA-like formula for the physical
electromagnetic response function $ {\hat \chi}^{-1}_{ij}(q)={\hat \Pi}^{-1}_{ij}(q)+2\pi{\hat
K}\epsilon_{ijk}{q^k/q^2}$ where $q^k=(\omega,{\bf q})$. Quantized values
of the Hall conductivity corresponding to the putative FQHE plateaus are given by the formula \be
\sigma_{xy}=\sum_\alpha^N\sigma_\alpha^{H}-\sum_{\alpha,\beta}^N\sigma_\alpha^{H} ({\hat \sigma}^H+{\hat
K}^{-1})^{-1}_{\alpha\beta}\sigma_\beta^{H} \ee where ${\hat \sigma}^H=(2\pi/\omega) Im{\hat
\Pi}_{xy}|_{\omega,{\bf q}\to 0}$ is a tensor of the CDF Hall conductivities.

As one important example of this general construction, attaching two units of the $\alpha$-type flux ($\int_{\bf
r}<{\bf \nabla}\times{\bf a}_\alpha>=\pm 4\pi$) to the CDFs of the same type is equivalent to choosing ${\hat K}=\pm
diag[2,\dots,2]$, in which case Eq.(3) yields \be \sigma_{xy}=\sum^N_\alpha{\nu_\alpha\over 2\nu_\alpha\pm 1}
\ee Notably, the overall Hall conductivity (6) is given by a "parallel" combination of the conductivities of the
individual species (each of which is, in turn, given by a "series" combination of the responses to the physical
electromagnetic $\bf A$ and the corresponding statistical field ${\bf a}_\alpha$). This composition rule should
be contrasted against the naive one, $\sigma_{xy}=\sum^N_\alpha\nu_\alpha/(2\sum_\alpha^N\nu_\alpha+1)$ (see,
e.g., the last reference in Ref.\cite{th}), which would have resulted from a series connection between the
response to $\bf A$ and a parallel combination of all the statistical fields (or a single field
that couples symmetrically to all the fermion species).

A more general case of the diagonal matrix ${\hat K}=diag[2p_1,\dots, 2p_N]$ gives rise to a formula
similar to Eq.(6) (with the factor of $2$ replaced by $2p_\alpha$ in the denominator of the $\alpha$-term).
Furthermore, any "entangled" way of attaching the fluxes described by a non-diagonal $\hat K$-matrix yields
an expression different from Eq.(6). Such alternative choices of the $\hat K$-matrix (which we do
not consider in this work) would be physically appropriate if different CDF species formed mutually coherent
states. In the spin-polarized ($N=2$) case, one example of this sort is provided by the matrix ${\hat
K}=\pmatrix{0 & 2\cr 2 & 0}$ which has been previously discussed in the context of the double-layer $\nu=1/2$
system.

The $SU(4)$-symmetry of Eq.(1) gets lowered in the presence of various symmetry-breaking terms of the form
$\delta L=\sum_{\alpha,\beta}^N\int_{\bf r}\Psi^\dagger_\alpha{\Lambda}_{\alpha\beta}\Psi_\beta$ which can be of both, single-particle and many-body, nature (observe 
that these terms retain their form after the statistical transformation, if the matrix $\hat \Lambda$ is diagonal).

At the mean-field (Hartree-Fock) level, the list of such
terms includes the (exchange-enhanced) Zeeman term (${\hat \Lambda}_Z= E_Z{\hat {\bf 1}}\otimes {\hat {\bf
1}}\otimes{\hat \sigma}_z$) and a parity-odd  mass term (${\hat \Lambda}_M=\Delta{\hat \sigma}_z\otimes{\hat
{\bf 1}}\otimes{\hat {\bf 1}}$ or $\Delta{\hat \sigma}_z\otimes{\hat {\bf 1}}\otimes{\hat \sigma}_z$), where the
first, second, and third factors refer to the sublattice, valley, and spin subspaces, correspondingly.
These two terms lift, respectively, the spin and valley degeneracies of the $0^{th}$ LL, thereby splitting it
into four individual sub-levels. Notice that the parity-even mass terms ($\sim{\hat \sigma}_z\otimes{\hat
{\sigma}}_z\otimes{\hat {\bf 1}}$ or ${\hat \sigma}_z\otimes{\hat \sigma}_z\otimes{\hat \sigma}_z$) which have
been previously discussed in the context of spin-orbit coupling and Spin Hall Effect in graphene can only split
the $0^{th}$ LL into $two$ sub-levels (with the Zeeman term present).

The symmetry-breaking terms appear to be instrumental for describing, e.g., the
plateau transitions $0\to \pm 1$, in which case all the four sub-levels of the $0^{th}$
LL are fully spin- and valley-resolved, as suggested by the strong-field ($B\gtrsim 20 T$) data of Ref.\cite{kim}.

The nearby FQHE states at $|\nu|<1$ can then be constructed with the use of single-component ($N=1$) CDFs which
occupy an integer number of the effective LLs. Naturally, these states fall into the standard Jain's series
converging towards $\nu=\pm 1/2$ \be \sigma^{N=1}_{xy}=\pm\nu^{\pm}_m=\pm{m\over 2m\pm 1} \ee where
$m=1,2,\dots$ and the overall $\pm$ sign is not correlated with that in the definition of $\nu^{\pm}_m$. Similar
fractions can occur near $\nu=\pm 3/2$, thereby giving rise to the FQHE plateaus at $\sigma_{xy}=\pm
(1+\nu^{\pm}_m)$. Since the single-particle CDF states are non-degenerate, a relative stability of the even vs
odd-numerator fractions (7) is not an issue (cf. with the discussion in the last of Ref.\cite{fqhe}).

In the case of a residual $SU(2)$ degeneracy of either spin or valley origin, the number of relevant CDF species
becomes $N=2$. Conceivably, such a situation can occur at the $0\to \pm 2$ plateau transitions (where, say,
$\nu_{L,R}^{\uparrow}=1/2$ or $\nu_{L}^{\uparrow,\downarrow}=1/2$, depending on the relative magnitude of $E_Z$
and $\Delta$).

The data of Ref.\cite{kim} suggest that at moderately strong fields ($10 T\lesssim B\lesssim 20 T$) the spin degeneracy gets lifted first (at
least, at the $n=\pm 1$ LLs). In this scenario, the residual valley degeneracy gives rise to a series of
valley-unpolarized IQHE states of the $N=2$ CDFs which converge towards $\nu=\pm 1$ and correspond to the
plateaus \be \sigma^{N=2}_{xy}=\pm (1-\nu^{\mp}_{m\pm 1}+\nu^{\pm}_{m})=\pm{2m\over {2m\pm 1}} \ee where
$\nu^{\pm}_m$ is defined in Eq.(7).

In Eq.(8), we took into account the fact that the numbers of occupied (spin-polarized) effective LLs for the
$L-$ and $R-$type CDFs differ by $one$ as a result of the spectral anomaly at the $0^{th}$ CDF LL (the $R$-type states reside at the energy $E=\Delta$, whereas the $L$-type ones are at $E=-\Delta$). 
As a result, the partial Hall
conductivities of the $R$- and $L$-species as functions of chemical potential obey the relation
$\sigma^R_{xy}(-\mu)=-\sigma^L_{xy}(\mu)$, although their sum $\sigma^R_{xy}+\sigma^L_{xy}$ is, of course, an odd function of $\mu$. It is worth noting that, from a formal standpoint, the anomalous IQHE observed in
Refs.\cite{exp} has the very same origin.

Lastly, $SU(4)$-invariant spin- and valley-unpolarized states
would be described in terms of $N=4$
CDFs which provide a mean-field picture of the $-2\to 2$ plateau transition
in terms of the half-filled $0^{th}$ LL
($\nu_{L,R}^{\uparrow,\downarrow}=1/2$) which is appropriate at relatively
weak fields ($B\lesssim 10 T$), according to the data of Ref.\cite{kim}.

Incompressible spin- and valley-unpolarized $N=4$ CDF states would then correspond to the plateaus \be
\sigma^{N=4}_{xy}=2(\nu^{\pm}_{m}-\nu^{\mp}_{m\pm 1})=\pm{2\over 2m\pm 1} \ee Notably, the series (9) includes
(pseudo)spin-singlet states at $\nu=2/3$ and $2/5$, thus providing a possible CDF picture of the exact ground
states found at these filling factors in the recent numerical studies \cite{fqhe}.

By analogy with the conventional 2DEG \cite{hlr}, we conjecture that the parent CDF states at
$\nu^{(N=1)}=k-1/2$, ($k=-1,0,1,2$), $\nu^{(N=2)}=\pm 1$ and $\nu^{(N=4)}=0$ for $N=1,2$, and $4$, respectively,
behave as compressible "CDF metals" characterized by the presence of a Fermi surface of radius $k^*_F=(2\nu B/N)^{1/2}$.

The mean-field CDF dispersion relation remains linear and the effective CDF velocity determined
by the strength of the Coulomb interaction, $v^*_{F}\sim gv_FN^{1/2}$, is comparable to $v_F$ for $g$ and $N$ of order one.
Due to their inherited Dirac kinematics, the Subnikov-de-Haas oscillations of the CDF resistivity 
at small deviations from the compressible fractions $\nu^{(N)}$ are expected 
to show the same Berry phase of $\pi$ as that of the original Dirac quasiparticles in weak fields \cite{exp}.

The CDF Fermi energy $E^*_{F}=v^*_{F}k^*_F\sim gv_FB^{1/2}$ appears to be of the same order as the distance
$E_1-E_0=v_F(2B)^{1/2}$ between the $0^{th}$ and $\pm 1^{th}$ LLs, suggesting that in graphene the LL mixing
effects are potentially more important than in the conventional 2DEG where they get suppressed with
increasing field. Moreover, the LL mixing becomes stronger with an increasing number $n$ of the occupied
electronic LLs, as the distance between the adjacent levels decreases as $\sim |n|^{-1/2}$. It is, therefore,
likely that most favorable for the formation of the parent CDF metals and their incompressible descendants is
the $0^{th}$ LL (cf. with the conclusions drawn in Refs.\cite{fqhe} where the LL mixing was neglected from the
outset).

In the CDF IQHE states (7,8,9), the energy gaps for well separated particle-hole excitations  \be
\Delta_m\approx E^*_{m}-E^*_{m-1}={v^*_F}(2B)^{1/2} {m^{1/2}-(m-1)^{1/2}\over (2m-1)^{1/2}}\ee  scale as
$\sim\sqrt B/m$ for large $m$, which dependence is similar to that found in the conventional case of
"non-relativistic" composite fermions where the effective mass varies as $\sim B^{1/2}$ (see Ref.\cite{hlr}). In
contrast, for small $m$ the true lowest energy excitations are likely to be represented by pairs of spin/valley
(anti)skyrmions \cite{fisher,fqhe}.

Despite the general possibility for the compressible states to emerge at any of the aforementioned fractions
$\nu^{(N)}$, a relative stability of these states is strongly dependent on the number $N$
of the CDF species involved. In order to proceed with the stability analysis one has to go beyond the
mean-field picture by including fluctuations of the statistical fields ${\bf a}_\alpha$ controlled by
the CDF polarization operator.

In the regime where typical CDF energies and momenta are small compared to $k^*_F$ and $E^*_F$,
respectively, ${\hat \Pi}_{ij}(q)$ is similar to that of a "non-relativistic" system with the same $k_F$ and
$v_F$. In particular, its transverse (with
respect to the transferred momentum $\bf q$) component
$\Pi^\perp(\omega,{\bf q})=
{\bf q}\times{\bar {\bf \Pi}}\times{\bf q}/{\bf q}^2= aq^2+ib\omega/q$, where $a\sim
v^*_F/k^*_F$ and $b\sim k^*_F$, accounts for the Landau diamagnetism and damping in the CDF metal.

For $N>1$ the CDF interactions are dominated by
$N-1$ linear combinations of the transverse components of the statistical fields which are orthogonal to the
"in-phase" mode $\sum^N_\alpha a_\alpha^i$. Unlike the latter, these combinations are not affected by the
unscreened Coulomb interactions, and the effective coupling between different CDF species (here $V_q=g/q$,
$({\bf v}_\perp{\bf v}_\perp^\prime)=({\bf v}{\bf v}^\prime)- ({\bf v}{\bf q})({\bf v}^\prime{\bf q})/{\bf
q}^2$)
\be
U_{\alpha\beta}=({\bf v}_\perp{\bf v}_\perp^\prime) {q^2V_q\over (Nq^2V_q+\Pi^\perp)\Pi^\perp}\approx ({\bf
v}_\perp{\bf v}_\perp^\prime){1\over N\Pi^\perp} \ee is always attractive in the Cooper channel
(${\bf v}=-{\bf v'}$). For $N=2$ this interaction can facilitate the onset of $s$-wave valley-singlet
pairing \cite{nick}. Moreover, for $N=4$ there exists a possibility of more exotic (spin-valley coupled)
patterns of the $SU(4)$-symmetry breaking.

By contrast, for $N=1$ the effective interaction is repulsive in the Cooper channel, as it is between any CDF
species of the same kind for $N>1$, \be U_{\alpha\alpha}=-({\bf v}_\perp{\bf v}_\perp^\prime)
{(N-1)q^2V_q+\Pi^\perp\over (Nq^2V_q+\Pi^\perp)\Pi^\perp}\approx -({\bf v}_\perp{\bf v}_\perp^\prime) {N-1\over
N\Pi^\perp} \ee Although there is still a possibility of $p$-wave pairing between the like CDFs, this potential
instability (which is also present for the $\nu=1/2$ state in the conventional 2DEG) tends to be much weaker
\cite{gww}.

Since the inherent pairing instabilities make the $N>1$ CDF metals prone to becoming incompressible paired
states, it is conceivable that the compressible states at the filling factors $\nu^{(2,4)}$ should, in general, be
less robust than those at $\nu^{(1)}$. Likewise, the chances of observing
the novel series (8) and (9) might be rather limited, as compared to the standard one given by Eq.(7).

The CDF metals would also be highly sensitive to disorder. In the presence of potential (short-range) impurities
of density $\rho_i$, the CDFs experience elastic scattering off of an effective random magnetic field whose
vector potential is described by the Gaussian variance $<A_i({\bf q})A_j({\bf
-q})>=16\pi^2\rho_i(\delta_{ij}-q_iq_j/{\bf q}^2)/{\bf q}^2$ \cite{hlr}.

A transport rate for the CDF $\alpha$-species can be estimated as $\Gamma^*\sim E^*_F\rho_i/|\rho_\alpha|$, and
the above analysis (including the role of the symmetry-breaking terms) pertains to the regime where $E_Z,
\Delta, v^*_FB^{1/2}/|m| \gtrsim \Gamma^*$. By the same token, disorder makes it more difficult to resolve
metallic states at fractions $\nu\sim 1/2p$ with $p>1$.

Evaluating the longitudinal conductivity of the CDF $\alpha$-species as $\sigma^{*}_{xx}\approx max [|\rho_\alpha|/\rho_i, 1]$, one
obtains a rough estimate for the physical conductivity of the CDF metals \be \sigma_{xx}\approx {1\over 4} min[\sum^N_\alpha{\rho_i\over |\rho_\alpha|},{N}] \ee which dependence
should be contrasted with that at zero field (in the case of Coulomb impurities, the latter is $proportional$ to
the total electron density $\sum_\alpha^N\rho_\alpha$ \cite{exp}). Interestingly enough, in the experiment of
Ref.\cite{kim} the conductivity at the $\nu=0$ plateau was found to be of order $\sigma_{xx}\approx 0.6$,
possibly suggesting a precursor of the formation of the $N=4$ CDF metal at weak fields.

Also, by analogy with the situation in the conventional 2DEG \cite{mirlin}, we predict that the conductivity of
the CDF metals is going to be temperature dependent due to quantum interference corrections which dominate over weak-(anti)localization ones and behave as
\be \delta\sigma_{xx}\propto
-\ln\sigma^*_{xx}\ln{\Gamma\over T}~~~ or~~~ -\ln^2{\Gamma\over T} \ee for $N=1$ and $N>1$, respectively,
thus allowing one to discriminate (in principal) between the single- and multi-component CDF metals.

Furthermore, despite the ostensibly Fermi-liquid-like properties of the CDF metals, the electron spectral
function $Im G({\bf p}, \epsilon)$ exhibits a distinctly non-Fermi-liquid behavior. Repeating the calculations
carried out in the case of the conventional 2DEG \cite{platzman}, we find a tunneling $I-V$ characteristics of
the CDF metal
 \be I(V)\propto \exp[-{\it Const}(E^*_F/V)^\eta] \ee where $\eta=1$ for $N=1$
and $1/2$ for $N>1$.

To conclude, in addition to a wealth of its other remarkable properties, graphene provides a natural venue for
the merger between the notions of pseudo-relativistic quasiparticles and statistical transformation from
electrons to CDFs. We predict that compressible CDF states are most likely to be observed at the $|\Delta\nu|=1$
plateau transitions (e.g., $0\to 1$) between the fully resolved sub-levels of the $0^{th}$ LL, since a stronger LL
mixing makes it more difficult for such states to form at the $|n|\neq 0$ LLs.

In contrast, the would-be CDF metals associated with the $|\Delta\nu|=2,4$ transitions are generally more
fragile due to their propensity towards pairing, which can drive these states incompressible. Moreover,
we predict that the incompressible CDF IQHE states can occur at both, the standard (7) as well as novel
(8,9), fractions.

As far as the practical possibility of testing these predictions is concerned, the detrimental effect of
disorder calls for performing experiments of Refs.\cite{exp,kim} in samples of substantially higher mobility.
The anticipated experimental signatures of the CDF metals can then be probed by such well-established techniques
as bulk tunneling, acoustic wave propagation, magnetic focusing, and other geometric resonances \cite{hlr}.

For one, if a compressible $\nu=0$ state were indeed to occur at $B\lesssim 10 T$, its CDF excitations would be amenable to
conventional electrostatic gating. This prediction should be contrasted with such a hallmark of the zero-field Dirac kinematics as the celebrated Klein's paradox that would hinder any possibility of electrostatic confinement of the electronic Dirac excitations with vanishing Fermi momentum \cite{exp}.

This research was supported by NSF under Grant DMR-0349881. The
author acknowledges valuable communications with
V.P. Gusynin, A. Geim, P. Kim, J. Smet, and the
hospitality at the Aspen Center for Physics
where this work was completed.

\end{document}